\documentclass[twocolumn,pra,floatfix,showpacs]{revtex4-1}
\usepackage{amsmath,amssymb,graphicx,epstopdf,epsfig}
\newcommand{\abs}[1]{\left|#1\right|}
\newcommand{\eva}[1]{\left<#1\right>}
\newcommand{\inleva}[1]{\langle#1\rangle}
\newcommand{\threevec}[3]{\left(\begin{array}{c}#1\\#2\\#3\end{array}\right)}
\newcommand{\nematic}{\mathbf{\hat{d}}}
\newcommand{\absF}{|\langle\mathbf{\hat{F}}\rangle|}
\newcommand{\SO}{\mathrm{SO}}
\newcommand{\U}{\mathrm{U}}

\newcommand{\eq}[1]{Eq.~(\ref{#1})}

\usepackage[usenames,dvipsnames]{color}
\usepackage[normalem]{ulem}

\begin{document}

\author{Justin Lovegrove}
\author{Magnus O.\ Borgh}
\author{Janne Ruostekoski}
%\email{janne@soton.ac.uk}
\affiliation{School of Mathematics, University of Southampton,
  Southampton, SO17 1BJ, United Kingdom}

\title{Energetically stable singular vortex cores in an atomic spin-1 Bose-Einstein condensate}

\begin{abstract}
We analyze the structure and stability of singular singly quantized vortices in a rotating spin-1
Bose-Einstein condensate. We show that the singular vortex can be energetically stable in both the
ferromagnetic and polar phases despite the existence of a lower-energy nonsingular coreless vortex
in the ferromagnetic phase. The spin-1 system exhibits an energetic hierarchy of length scales resulting
from different interaction strengths and we find that the vortex cores deform to a larger size
determined by the characteristic length scale of the spin-dependent interaction.
We show that in the ferromagnetic phase the resulting stable core structure, despite apparent complexity, can
be identified as a single polar core with axially symmetric density profile which is nonvanishing everywhere.
In the polar phase, the energetically favored core deformation leads to a splitting of a singly quantized
vortex into a pair of half-quantum vortices that preserves the topology of the vortex outside the extended core region, but breaks the axial symmetry of the core. The resulting half-quantum vortices exhibit nonvanishing ferromagnetic cores.
\end{abstract}
\date{\today}
\maketitle

\section{Introduction}
\label{sec:introduction}

In the textbook examples of superfluids, liquid $^4$He~\cite{tilley} and
Bose-Einstein condensates (BECs)~\cite{pethick-smith} of spinless or spin-polarized atoms, quantized vortices
occur as quantized circulation
around an empty vortex core whose size is determined by a
characteristic healing length.
In a BEC of atoms whose spin degree of freedom is not frozen by
magnetic fields~\cite{stenger_nature_1998} spin rotations and condensate phase combine to form a larger set of physically distinguishable degenerate states. This is analogous to
liquid $^3$He where superfluidity is formed by Cooper pairs of fermions that exhibit a nonzero spin and orbital angular momentum, resulting in a rich phenomenology of phases with different broken symmetries~\cite{vollhardt-wolfle}. A variety of different vortex configurations~\cite{salomaa_rmp_1987} and other defects and textures~\cite{volovik} have been theoretically studied and experimentally observed in the resulting multi-component order parameter manifolds of superfluid liquid $^3$He. There are obvious parallels~\cite{volovik} to similar objects in cosmology~\cite{vilenkin-shellard} and quantum field theory~\cite{manton_sutcliffe}.

Consequently, it is not surprising that in multicomponent BECs, there has been an increasing interest in theoretical studies of topological defects and textures~\cite{ho_prl_1998,yip_prl_1999,leonhardt_jetplett_2000,ruostekoski_prl_2001,alkhawaja_nat_2001,
stoof_monopoles_2001, isoshima_pra_2002,
mizushima_pra_2002,mizushima_prl_2002,martikainen_pra_2002,
mueller_prl_2002,kasamatsu_prl_2003, battye_prl_2002, savage_prl_2003,
ruostekoski_monopole_2003,zhou_ijmpb_2003,savage_dirac_2003,reijnders_pra_2004,
ruostekoski_pra_2004,mueller_pra_2004,saito_prl_2006,santos_spin-3_2006,semenoff_prl_2007,barnett_pra_2007,pietila_prl_2009,huhtamaki_pra_2009,
kobayashi_prl_2009,takahashi_pra_2009,simula_jpsj_2011,borgh_interface_2012,
nitta_arxiv_2012} as well as vector
solitons~\cite{busch_prl_2001,ohberg_prl_2001,kevrekidis_epjd_2004,berloff_prl_2005,shrestha_prl_2009,yin_pra_2011,nistazakis_pra_2008,carretero-gonzales_nlin_2008}.
The theoretical work has been produced parallel with a rapid experimental progress on spinor BECs, e.g., in controlled preparation of coreless spinor vortices~\cite{leanhardt_prl_2003,leslie_prl_2009,choi_prl_2012}, in the studies of spin-texture formation~\cite{vengalattore_prl_2008,kronjager_prl_2010}, and in nonequilibrium vortex production during rapid phase transitions~\cite{sadler_nature_2006}.

An atomic spin-1 BEC exhibits two phases of the ground-state manifold, ferromagnetic (FM) and polar,
with distinct broken symmetries. In the FM phase the ground state of a sufficiently rapidly rotating
atom cloud is formed by nonsingular, coreless vortices in which the order parameter is well-defined
everywhere~\cite{mizushima_prl_2002,martikainen_pra_2002,reijnders_pra_2004,mueller_pra_2004,takahashi_pra_2009}. Similar coreless vortices were first described in superfluid liquid $^3$He~\cite{anderson_prl_1977,mermin_prl_1976} and were recently experimentally phase-imprinted on a spinor BEC~\cite{leslie_prl_2009}. Due to the topology of the FM ground-state manifold, which is defined by the group of spin rotations, it is also possible to form a singular vortex~\cite{isoshima_pra_2002,mizushima_pra_2002}, whose stability and structure, however, is much less well understood.

\begin{figure}[tb]
  \centering
  \includegraphics[width=0.9\columnwidth]{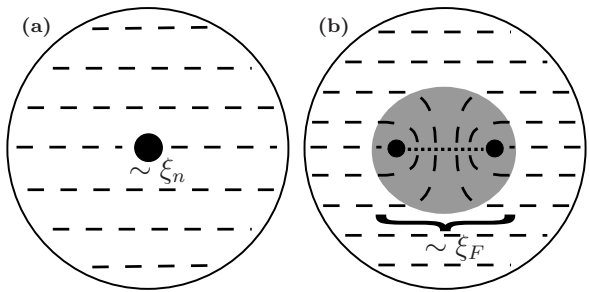}
  \caption{Schematic illustration of two vortex-core structures with
    the same topology for a singly quantized singular vortex in the
    polar phase of a spin-1 condensate. In (a) the atom density vanishes
    at the vortex-line singularity with the core size determined by the
    characteristic length scale $\xi_n$ (healing length) associated with
    the spin-independent interaction strength. In (b) the atom density
    is nonvanishing in the core region whose size is determined by the
    characteristic length scale $\xi_F$ of the spin-dependent
    interaction strength. The vortex line singularity has now split into
    two half-quantum vortices with the atoms in the ferromagnetic phase
    at the precise location of the singularities. In both figures we
    show the nematic axis as a dashed line and the dotted line in (b)
    indicates a disclination plane for the nematic axis. Inside the core
    region (shaded area) of (b) the broken symmetry of the polar ground
    state manifold is restored (as explained in the text). Outside the core
    the topological properties of the vortex are the same as those in
    (a).
  }
  \label{fig:core-splitting-schematic}
\end{figure}

Here we show that singular, singly quantized vortices can be
energetically stable in both FM and polar phases of a spin-1 BEC. In the FM phase this is despite the
fact that the coreless vortex has a lower energy.  Although a singular
vortex would also not be nucleated by rotation, once created, for
example by phase-imprinting, it can be stabilized in a rotating trap. In the polar phase, the singular vortex undergoes a core deformation to a pair of half-quantum vortices in an extended vortex core region where the broken order parameter symmetry of the polar ground-state manifold is restored (see Fig.~\ref{fig:core-splitting-schematic}).

In a singular defect the singularity of the order parameter is contained by a
defect core.  Unlike scalar superfluids, in the spinor BEC this
does not imply that the density must vanish: it is also possible to
accommodate the singularity by requiring the spinor wavefunction to be orthogonal to the ground-state manifold at the precise location of the singularity.
The different possibilities for the defect core structure leads to an energetic hierarchy of different characteristic length scales~\cite{ruostekoski_monopole_2003}: Depending on the ratio of the spin $\xi_F$ and density $\xi_n$ healing lengths associated with the two interaction strengths of spin-1 BEC, it can be
energetically more favorable to force the order parameter value to be orthogonal to the ground-state
manifold at the defect singularity than to force the density to
zero. This can lead to unexpected core structures. In
Ref.~\cite{ruostekoski_monopole_2003} it was shown how in the polar
phase of a spin-1 BEC a singular point defect with a vanishing density
at the singularity can spontaneously deform to a ring defect with a FM
core and a nonvanishing density.
Recently symmetry classification using homotopy theory was used in the analysis of defect cores in Ref.~\cite{kobayashi_arxiv_2012}.

In a spin-1 BEC the polar and FM phases differ by the local
expectation value of the spin magnitude.
The size of the filled vortex core is then determined by $\xi_F$ that
defines the length scale over which the spin magnitude heals when
locally perturbed. This is in general much larger than the size
$\xi_n$ of a density-depleted vortex core. Large wavefunction
gradients close to the defect singularity result in a large
order-parameter bending energy.  Energetically, the system
therefore prefers the larger core size and a nonvanishing atom density
with correspondingly lower bending energy.
Outside the filled core region of size $\xi_F$  the
topology of the vortex is the same as in the case of a zero-density
vortex line. In that region the order parameter bending energy is not
sufficient to excite the system away from the ground-state manifold
and we find a well-defined broken order parameter symmetry of either
the polar or FM phase. It is only inside the filled core of size
$\xi_F$ where the vortex structure differs. Inside the core the order
parameter bending energy restores the order parameter symmetry of the
full spin-1 condensate wavefunction by exciting the system out of the
ground-state manifold by mixing the polar and FM phases. In our
numerical simulations this is indicated by a continuously varying spin
magnitude across the vortex core.

We show that in the case of a singular singly quantized FM vortex the apparent complexity
of the core can in this way be explained as the formation of a
  single core with $\absF=0$ at the singularity:
An initial singular vortex is formed by
overlapping vortex lines in the spinor components. As the system
relaxes, the density depletion is avoided by separating the vortex
lines. By a rotation of the spinor basis, the axial symmetry of density profiles
of the individual spinor components with perfectly
  overlapping vortex lines is explicitly restored,
and the vortex is identified as one in
which the spin vector winds by $2\pi$ around a polar core.
We find both axisymmetric and nonaxisymmetric solutions for the singular vortex
in the FM phase, indicating close energetic degeneracy of the solutions.

In the polar phase, we use the same analysis to show how it is energetically favorable
for the system to spontaneously break axial symmetry by splitting the
core of a singly quantized vortex into two FM cores.  The resulting
spinor wavefunction shows a complex combination of vortex lines with highly deformed anisotropic cores.  However by transforming the spinor wavefunction to the basis specified by the direction of
the spin in the FM cores, the spinor
structure is identified as a pair of half-quantum vortices.
Outside the deformed core, the topology of the initial
singly quantized vortex is preserved.
The splitting can be understood from the nematic symmetry properties of the polar order parameter.
A stable nonaxisymmetric singular vortex with a nonzero superfluid density at the core has been theoretically predicted~\cite{thuneberg_prl_1986,salomaa_prl_1986} and
experimentally observed~\cite{kondo_prl_1991} in superfluid liquid $^3$He.

Here we analyze the energetic stability of the singular vortices and explain
the structures of the their cores by numerical simulations in the
framework of mean-field theory. The paper is organized as follows: In
Sec.~\ref{sec:mft} we give a brief overview of mean-field theory for
the spin-1 BEC and explain the general concepts used in our
analysis. In Sec.~\ref{sec:fm-vortex} we demonstrate the energetic stability
and explain the core structure of a singular vortex in the FM phase.
Section~\ref{sec:polar-vortex} applies a similar analysis to study
energetic stability and identify the core-structure of a singular vortex
in the polar phase.  We conclude with a brief summary of our findings
in Sec.~\ref{sec:conclusions}. Analytic properties of the vortex solutions and the
basis transformations are provided in Appendix~\ref{append}.

\section{Spin-1 mean-field theory}
\label{sec:mft}

In our analysis of singular vortices of a spin-1 atomic BEC, we consider the classical (Gross-Pitaevskii) mean-field theory of a harmonically trapped system that results in a spatially nonuniform atom density.
In an optical trapping potential the atomic spin is not frozen by magnetic fields and the spin-1 BEC
is represented by a normalized three-component spinor $\zeta(\mathbf{r})$ in the basis of spin
projection onto the $z$ axis. Together with the density
$n(\mathbf{r}) = |\Psi(\mathbf{r})|^2$ this specifies the macroscopic condensate wavefunction
\begin{equation}
  \label{eq:spinor_suppl}
  \Psi({\bf r}) = \sqrt{n({\bf r})}\zeta({\bf r})
  = \sqrt{n({\bf r})}\threevec{\zeta_+({\bf r})}
                              {\zeta_0({\bf r})}
                              {\zeta_{-}({\bf r})},
  \quad
  \zeta^\dagger\zeta=1.
\end{equation}
The Hamiltonian density in the frame rotating with frequency $\Omega$
around the $z$ axis
is~\cite{ho_prl_1998,ohmi_jpsj_1998,pethick-smith}
\begin{equation}
  \label{eq:hamiltonian-density}
  \begin{split}
    {\cal H} &=  \frac{\hbar^2}{2m}\abs{\nabla\Psi}^2 + V(\mathbf{r})n
    + \frac{c_0}{2}n^2
    + \frac{c_2}{2}n^2\abs{\mathbf{\eva{\hat{F}}}}^2\\
    &+ g_1n\eva{\mathbf{B}\cdot\mathbf{\hat{F}}}
    + g_2n\eva{\left(\mathbf{B}\cdot\mathbf{\hat{F}}\right)^2}
    -\Omega\eva{\hat{L}_z},
  \end{split}
\end{equation}
where $V(\mathbf{r})$ is an external trapping potential, $m$ is the atomic
mass and
$\langle\hat{L}_z\rangle=-i\hbar\Psi^\dagger(x\partial_y-y\partial_x)\Psi$
denotes the
$z$-component of the angular momentum
operator.  The spin operator
$\mathbf{\hat{F}}$, whose expectation value
$\mathbf{\inleva{\hat{F}}}=\zeta_\alpha^\dagger\mathbf{\hat{F}}_{\alpha\beta}\zeta_\beta$
yields the local spin vector, is given by
a vector of spin-1 Pauli matrices.
The first two terms in the second line of
Eq.~(\ref{eq:hamiltonian-density}) describe linear and quadratic Zeeman shifts,
respectively, in the presence of a weak external magnetic field
$\mathbf{B}$.
In this paper we assume an axially symmetric harmonic confinement such
that
\begin{equation}
  V(\mathbf{r})
  =\frac{1}{2}m\left[\omega_\perp^2 (x^2+y^2)+\omega_z^2 z^2\right],
\end{equation}
from which we also define the transversal oscillator length $l_\perp =
\sqrt{\hbar/m\omega_\perp}$.

The spins of two colliding spin-1 atoms may combine to a relative
angular momentum of either $0$ or $2$.  This implies that the contact
interaction results from two different contributions, corresponding
to the
two scattering channels with different
$s$-wave scattering lengths $a_0$ and $a_2$.  The two scattering contributions lead to the two
interaction terms in Eq.~(\ref{eq:hamiltonian-density}). The strength of the interactions may be
calculated using angular momentum algebra and we have
$c_0=4\pi\hbar^2(2a_2+a_0)/3m$ and $c_2=4\pi\hbar^2(a_2-a_0)/3m$~\cite{pethick-smith}.
In addition the Hamiltonian density \eqref{eq:hamiltonian-density} may include magnetic dipole-dipole interaction terms that can influence the spin textures~\cite{vengalattore_prl_2008,lovegrove}.

The spin-dependent interaction term $c_2$ in \eq{eq:hamiltonian-density} determines the spin magnitude
in a uniform ground-state spin distribution.
If $c_2<0$, as is the case for $^{87}$Rb, the spin-dependent
contribution to the interaction energy
will favor the FM state with $\absF=1$ throughout the BEC.
Conversely if $c_2>0$, as for $^{23}$Na, the
polar state with $\absF=0$ will be favored.

The two interaction strengths $c_0$ and $c_2$ are each associated with
a characteristic length scale.
From the spin-independent interaction we can derive the healing length
$\xi_n=(8\pi c_0 n)^{-1/2}$ that defines the length scale over which
the density heals around a
local depletion of the atom density~\cite{pethick-smith}. This phenomenon
is similar in a scalar BEC,
which exhibits a healing length depending on the atom density and the
scattering length. Due to the spin-dependent interaction term, we now,
however, have an additional healing length, analogously
given by $\xi_F=(8\pi |c_2| n)^{-1/2}$.  This defines the length scale over which the spin magnitude
$|\langle\mathbf{\hat{F}}\rangle|$ heals when locally
perturbed.

As in a scalar BEC, single-valuedness of the order parameter may be
maintained at a defect singularity by requiring that the density vanishes
there. The size of the defect core is then given by the healing length
$\xi_n$.  However, the spinor order parameter makes it possible to
maintain a nonzero density at the cost of
requiring that the wavefunction at the singularity becomes orthogonal
to the ground-state manifold.  For example, a singularity in the FM
manifold where $\absF=1$ can be accommodated by having $\absF=0$ on
the vortex line.  This constitutes a local perturbation of the spin
magnitude, and so its length scale is determined by $\xi_F$, which is
usually larger than $\xi_n$.  The
energetic cost of the local change in spin magnitude due to increased
interaction energy may be offset by
the lower bending energy in the larger core.

The two phases of the spin-1 BEC are described by very different order
parameters, which leads to dramatically different possible vortex
states. In the following we shall first consider
the FM phase and show
that a singular, singly quantized vortex can be energetically stable,
and describe how its core structure can be understood in terms of the
energetics of characteristic length scales [Sec.~\ref{sec:fm-vortex}].
We will then apply a similar analysis to show how the deformed core of
a stable singly quantised vortex in the polar phase can be identified
as a pair of half-quantum vortices [Sec.~\ref{sec:polar-vortex}].

\section{Stability and core deformation of a singular ferromagnetic vortex}
\label{sec:fm-vortex}

We first consider vortices in the FM phase of a spin-1 BEC. The system becomes FM when the interaction term
$c_2<0$ in the Hamiltonian \eqref{eq:hamiltonian-density}; energetically it is then favorable to maximize
the spin magnitude everywhere in space, so that $\absF=1$. A general FM spinor wavefunction can be
constructed from the representative spinor $\zeta = (1,0,0)^T$ with
$\inleva{\mathbf{\hat{F}}}=\mathbf{\hat{z}}$ by incorporating a macroscopic condensate phase $\phi$ and by a
spin rotation $U(\alpha,\beta,\gamma)=\exp(-iF_z\alpha)\exp(-iF_y\beta)\exp(-iF_z\gamma)$,
defined by three Euler angles. We obtain
\begin{equation}
  \label{eq:fm}
  \begin{split}
    \zeta^{\rm f} &=
    e^{i\phi}U(\alpha,\beta,\gamma)\threevec{1}{0}{0}\\
    &= \frac{e^{-i\gamma^\prime}}{\sqrt{2}}
    \threevec{\sqrt{2}e^{-i\alpha}\cos^2\frac{\beta}{2}}
             {\sin\beta}
             {\sqrt{2}e^{i\alpha}\sin^2\frac{\beta}{2}}\,,
  \end{split}
\end{equation}
where \mbox{$\gamma^\prime=\gamma-\phi$}. The local spin vector is
then given by $\inleva{\mathbf{\hat{F}}}
=(\cos\alpha\sin\beta,\sin\alpha\sin\beta,\cos\beta)$.

Order-parameter space is the manifold of energetically degenerate
spinors $\zeta$.  Degenerate FM
spinors [Eq.~(\ref{eq:fm})] differ only by rotations in spin-space given
by the Euler angles $\alpha$, $\beta$ and $\gamma^\prime$.
The order-parameter space therefore corresponds to the group of
three-dimensional rotations $\SO(3)$.

\begin{figure}[tb]
  \centering
  \includegraphics[scale=1]{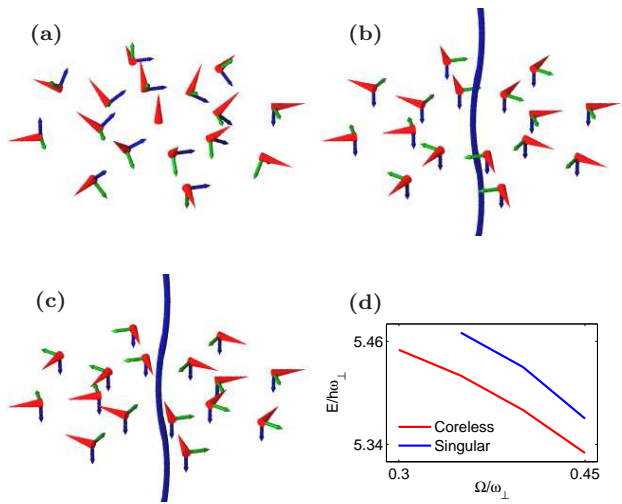}
  \caption{(color online) Schematic illustrations of FM vortex states.
    (a) The non-singular, coreless vortex is
    formed as a combined disgyration of the spin vector (cones) and a spin
    rotation about the
    local spin vector (indicated by the orthogonal
    vectors). The vortex is nonsingular and the spin texture is continuous.
    (b) The singular FM spin vortex is formed as a radially-oriented
    disgyration of
    the spin vector around the singular core, which is
    filled by the polar phase.
    (c) The class of singular vortices contain serveral different spin
    configurations, for example the cross-disgyration shown.  They can
    all be deformed to the singular spin vortex through local spin
    rotations.
    (d)  Energies of the stable coreless and
    singular vortices as functions of the frequency of rotation in an
    isotropic trap using $Nc_0=1000\hbar\omega_{\perp}l_{\perp}^3$ and
    $Nc_2=-320\hbar\omega_{\perp}l_{\perp}^3$. The coreless vortex is
    lower in energy in the whole frequency range.
  }
  \label{fig:fm-vortex-schematic}
\end{figure}

The topological stability of line defects is characterized by
the way closed contours encircling the defect map into order parameter
space~\cite{mermin_rmp_1979}. If the order-parameter space image of such a
closed loop can be contracted to a point, the defect is not
topologically stable. $\SO(3)$ may be
represented geometrically as $S^3$ (the unit
sphere in four dimensions) with diametrically opposite points
identified. The only closed loops that cannot be contracted to a point
are those connecting such identified points an odd number of times
(but these loops can all be deformed into one another).
There are therefore only two
distinct classes of vortices: singular vortices
corresponding to non-contractible loops, and non-singular vortices
corresponding to contractible loops~\cite{ho_prl_1998,ohmi_jpsj_1998}.
All the singular vortices with an odd-integer winding number are therefore topologically equivalent to a singly quantized singular vortex and all singular vortices with an even-integer winding number are topologically equivalent to a non-singular vortex-free state. Mathematically, this is indicated by the first homotopy group of $\SO(3)$ which has two elements [$\pi_1(\SO(3))=\mathbb{Z}_2$] that represent the two topological equivalence classes for the vortices.
Typical examples of a non-singular coreless vortex forming a
continuous spin texture and singular spin
vortex with radial and cross disgyrations of the spin vector are
schematically illustrated in Fig.~\ref{fig:fm-vortex-schematic}.

In the FM phase, circulation need not be
quantized~\cite{ho_prl_1998}.  A striking
manifestation of this fact is the possibility of having a coreless
vortex, which may be written
\begin{equation}
  \label{eq:cl}
  \zeta^{\rm cl} =
  \frac{1}{\sqrt{2}}\threevec{\sqrt{2}\cos^2\frac{\beta}{2}}
                             {e^{i\varphi}\sin\beta}
                             {\sqrt{2}e^{2i\varphi}\sin^2\frac{\beta}{2}},
\end{equation}
where $\varphi$ is the azimuthal angle and the Euler angle $\beta$
varies with the radial distance $\rho=\sqrt{x^2+y^2}$ from the
$z$ axis. The spin texture is kept continuous by having
$\beta\rightarrow0$ as $\rho\rightarrow0$.
The coreless vortex carries angular momentum, yet can be continuously
deformed into a vortex-free state.  Because the FM order parameter is
well-defined and non-singular everywhere in space (a coreless vortex
forms a continuous spin texture)
the coreless vortex belongs to the class of
non-singular vortices.  It is the spinor-BEC analogue of the
Anderson-Toulouse and Mermin-Ho vortices in
$^3$He~\cite{anderson_prl_1977,mermin_prl_1976},
which differ by
the imposed boundary conditions at the container wall.
The superfluid
velocity~\cite{ho_prl_1998}
\begin{equation}
  \mathbf{v} =
  \frac{\hbar}{m\rho}(1-\cos\beta)\boldsymbol{\hat{\varphi}}
\end{equation}
goes smoothly to zero at the center of the vortex but increases away
from it as $\beta$ increases. The spin forms a characteristic fountain
texture. In the atomic
gas, there is no hard container wall, and the amount by which
$\inleva{\mathbf{\hat{F}}}$ turns as $\rho$ increases from the vortex
center to the edge of the cloud is not fixed.  The total circulation
can thus vary smoothly as the value of $\beta$ at the edge of the
cloud adapts to the imposed rotation \footnote{In $^3$He the angular momentum
is fixed at the cylinder wall, fixing the boundary condition of the coreless
vortex texture. For a fixed boundary condition the system can exhibit a conserved
winding number defined by an integral of the spin vector expression similar to the
topological charge density of a point defect
over the surface covering the upper hemisphere~\cite{vollhardt-wolfle}.}.

The simplest way to construct a singly quantized singular vortex in
the FM phase is as
a $2\pi$-winding of the condensate phase $\phi$.  The vortex can then be described by the spinor
\begin{equation}
  \label{eq:fm-singular}
  \zeta^{\rm s} =
  \frac{e^{i\varphi}}{\sqrt{2}}\threevec{\sqrt{2}\cos^2\frac{\beta}{2}}
                                       {\sin\beta}
                                       {\sqrt{2}\sin^2\frac{\beta}{2}},
\end{equation}
where the density is required to vanish on the singular vortex line along the $z$ axis (where all
the three spinor components are singular).  The Euler angle $\beta$ is
arbitrary but constant, giving a uniform spin distribution (which,
without loss of generality, we assume to be in the $xz$ plane such
that $\alpha=0$ in Eq.~(\ref{eq:fm})).

We may continuously deform $\zeta^\mathrm{s}$ into another vortex in
the same class of topological line defects through purely
local operations. For example we may rotate the spins into the radial
disgyration ($2\pi$ rotation) of the spin vector
shown in Fig.~\ref{fig:fm-vortex-schematic}(b). This spin structure is
derived from Eq.~(\ref{eq:fm}) by letting $\alpha=\varphi$ while
$\gamma^\prime=0$,
\begin{equation}
  \label{eq:spinvortex}
  \zeta^{\rm sv} =
  \frac{1}{\sqrt{2}}\threevec{\sqrt{2}e^{-i\varphi}\cos^2\frac{\beta}{2}}
                             {\sin\beta}
                             {\sqrt{2}e^{i\varphi}\sin^2\frac{\beta}{2}},
\end{equation}
yielding a singular vortex with a circulation of the spin around the core (spin vortex).
This is similar to a radial disgyration of the angular momentum in an analogous vortex structure in $^3$He~\cite{vollhardt-wolfle}.
The FM order parameter is still singular at $\rho=0$ because a
singularity is introduced in the FM spin vector. At $\beta=\pi$ we recover the vortex in Eq.~(\ref{eq:fm-singular}) with only one spinor component occupied.
Further local rotations of the spin allow us to construct additional members
of the family of singular FM vortices.  If we
locally rotate all spins through $\pi/2$ around the $z$
axis in Fig.~\ref{fig:fm-vortex-schematic}(b) or
Eq.~(\ref{eq:spinvortex}), we change from the radial to a tangential
disgyration, where
$\inleva{\mathbf{\hat{F}}}=\boldsymbol{\hat{\varphi}}$. A spin vortex
could also be constructed from Eq~(\ref{eq:fm}) by choosing $\alpha=-\varphi$.
The radial disgyration is then replaced by the
cross disgyration illustrated in
Fig~\ref{fig:fm-vortex-schematic}(c).  Because the $\SO(3)$
order-parameter manifold allows only two topologically distinct
classes of vortices, all singly quantized, singular vortices can be
transformed into each other by local spin rotations and this family of vortices is indeed quite large.

In order to determine the energetic stability of the vortex configurations and stable
vortex core structures, we numerically minimize the energy of specific vortex states belonging to
distinct topological equivalence classes. The energy relaxation is done
by numerically propagating a coupled set of Gross-Pitaevskii equations derived
from Eq.~(\ref{eq:hamiltonian-density}) in imaginary time using a
split-step algorithm~\cite{javanainen_jpa_2006}. We consider an
isotropic trap in a rotating frame with the nonlinearities
$Nc_0=1000\hbar\omega_{\perp}l_{\perp}^3$ and
$-640\hbar\omega_{\perp}l_{\perp}^3 \leq Nc_2 \leq
-10\hbar\omega_\perp l_{\perp}^3$, where $N$ is the total number of
atoms.   As an initial state for a
singular singly quantized
vortex we take the vortex of Eq.~(\ref{eq:fm-singular}) in which case
each spinor component exhibits a singly quantized vortex line. These
all perfectly overlap with a vanishing density at the core.
We also perform an energy minimization of the coreless,
non-singular vortex of \eq{eq:cl}. The coreless vortices have been shown to
exist in the ground state of sufficiently rapidly rotating FM spin-1
BECs~\cite{mizushima_prl_2002,martikainen_pra_2002,reijnders_pra_2004,mueller_pra_2004},
and increasing the rotation rate of a vortex-free cloud is predicted
to result in nucleation of coreless vortices in the system.

\begin{figure}[tb]
 \centering
  \includegraphics[scale=1]{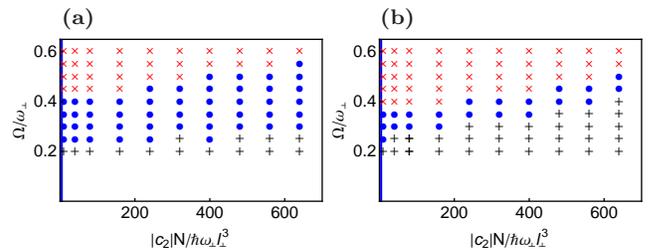}
  \caption{(color online) Energetic stability of the coreless (a) and the
    core-deformed singular
    vortex (b) in an isotropic trap for varying
    spin-dependent interaction strength $c_2<0$.  The spin-independent
    interaction is fixed at $Nc_0=1000\hbar\omega_{\perp}l_{\perp}^3$.
    Blue dots ($\bullet$) indicate that the vortex is energetically
    stable. A black plus ($+$) indicates where the initial vortex
    leaves the
    cloud, whereas red crosses ($\times$) mark where additional
    vortices nucleate
    due to rotation. A blue (black) vertical line marks
    $c_0/c_2\simeq-216$ relevant for
    $^{87}$Rb~\cite{van-kempen_prl_2002}. Note that with the
    parameters used here, this yields
    $N|c_2|=4.6\hbar\omega_{\perp}l_{\perp}^3$. The line thus falls very nearly on top of the vertical axis and the two cannot be distinguished in the figure.}
  \label{fig:isotropic-stability}
\end{figure}

\begin{figure}[tb]
  \centering
  \includegraphics[scale=1]{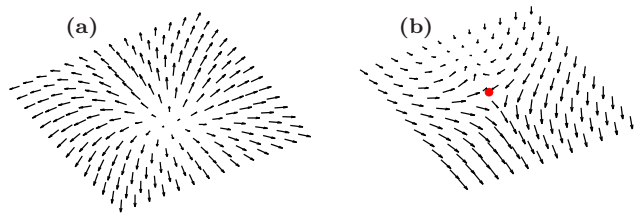}
  \caption{(color online) Numerically calculated spin textures in the
    stable FM vortex
    states in a rotating trap. The spin vector is shown in a cut
    perpendicular to the $z$ axis (the axis of rotation).
    (a) The spin vector in the coreless vortex exhibits a
    characteristic fountain-like structure and maintains $\absF=1$
    everywhere.
    (b) In the relaxed singular vortex, the spin vector winds by
    $2\pi$ around the $x$ axis on a path encircling the singular
    vortex core (indicated by the dot),
    in which $\absF \to 0$. This texture can be
    continuously deformed into that shown in
    Fig.~\ref{fig:fm-vortex-schematic}(b).
  }
  \label{fig:cl-v-singular}
\end{figure}

We find a single coreless vortex to be energetically stable in a
sufficiently rapidly rotating trap, as shown
in the stability diagram of
Fig.~\ref{fig:isotropic-stability}(a).
Figure~\ref{fig:cl-v-singular}(a) shows the characteristic
fountain-like spin texture of the stable vortex.  At
slow rotation speeds the vortex exits the atom cloud and at faster
rotation rates we observe nucleation of additional coreless vortices
to the system. The threshold rotation frequency is increased at
stronger nonlinearities. Our findings are consistent with those in
Ref.~\cite{martikainen_pra_2002}.

For the singular initial-state vortex the corresponding stability
diagram is displayed in
Fig.~\ref{fig:isotropic-stability}(b). Although its core structure is
deformed during energy relaxation (as we will discuss below), we find
that the singular vortex is energetically stable for a range of
rotation
frequencies at all investigated values of $c_2$. This energetic
stability of the singular vortex seems surprising since there also
exists a stable coreless vortex with lower energy at the
same rotation frequencies and nonlinearities. Our numerics also
show that coreless vortices will nucleate due to rotation, whereas
singular vortices will not. A comparison between the numerically
calculated energies of a stable coreless vortex and a stable singular
vortex as a function of the rotation frequency in an isotropic trap
and with
$Nc_0=1000\hbar\omega_{\perp}l_{\perp}^3$,
$Nc_2=-320\hbar\omega_{\perp}l_{\perp}^3$ is shown in
Fig.~\ref{fig:fm-vortex-schematic}(d). This may be contrasted with the
vortex energetics of coreless and singular vortices in superfluid liquid
$^3$He-$A$, where the singular vortex has lower
energy, but the energy barrier for nucleation of the singular core is
higher than that for forming a coreless vortex~\cite{parts_prl_1995}. Singular vortices
can be created by cooling a rotating normal fluid through the superfluid
transition.

The coreless and the singular vortices belong to distinct topological
equivalence classes and they cannot
be continuously deformed to each other. For the singular vortex to
decay, the rotation frequency has to be sufficiently slow so that the
vortex can exit the atom cloud and be replaced by a nucleating
coreless vortex that enters from the edge of the cloud. We find a
range of frequencies and nonlinearities
[Fig.~\ref{fig:isotropic-stability}(b)] for which the singular vortex
remains in the atom cloud and no additional coreless vortices
nucleate.
A single, singly quantized singular vortex thus represents a
\emph{local} minimum of the energy, topologically protected against
decay to the lower-energy coreless vortex.

\begin{figure}[tb]
  \centering
  \includegraphics[scale=1]{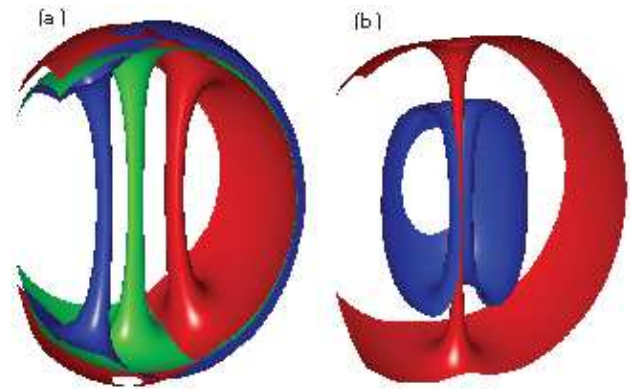}
  \caption{(color online) Split core of
    the singular FM vortex and restoration of a single core with
    explicit axial
    symmetry of the spinor-component densities
    illustrated by
    isosurfaces of the spinor wavefunction components
    $n\abs{\zeta_+}^2$ (red/medium gray), $n\abs{\zeta_0}^2$ (green/light gray) and
    $n\abs{\zeta_-}^2$ (blue/dark gray).
    (a) In the spinor basis along the $z$
    axis, the vortex lines in $\zeta^{(z)}_+$, $\zeta^{(z)}_0$ and
    $\zeta^{(z)}_-$ separate and the atom density is nonzero everywhere.
    (b) The axial symmetry of the density in each spinor component is restored by transforming the spinor to the basis of spin projection onto the x axis. Vortex lines
    with opposite circulation in $\zeta^{(x)}_\pm$ overlap.
    $\zeta^{(x)}_0$ (not shown) does not exhibit
    any vortex line (cf.\ Fig.~\ref{fig:split-core-1D}(b)). See also
    Appendix~\ref{append} for a qualitative analytic discussion of the
    relation between $\zeta^{(x)}$ and $\zeta^{(z)}$.}
  \label{fig:split-core-3D}
\end{figure}

\begin{figure}[tb]
  \centering
  \includegraphics[scale=1]{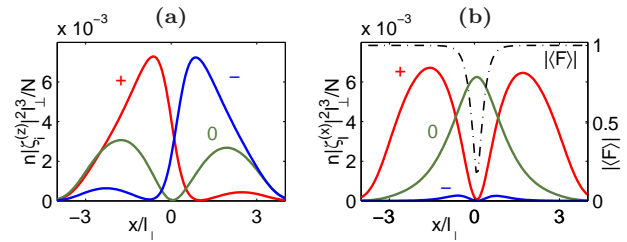}
  \caption{(color online) (a) Densities in the three spinor components
  $\zeta_+^{(z)}$ (red line marked by $+$), $\zeta_0^{(z)}$ (green line marked by $0$) and
  $\zeta_-^{(z)}$ (blue line marked by $-$) on the axis connecting the vortex
  lines in the spinor
  components (cf.\ Fig~\ref{fig:split-core-3D}(a)).
  (b) Densities in
  $\zeta_+^{(x)}$ (red line marked by $+$), $\zeta_0^{(x)}$ (green line marked by $0$)
  and $\zeta_-^{(x)}$ (blue line marked by $-$) on the same spatial axis after spinor
  basis transformation.  $\absF$ (black dash-dotted line) goes to zero
  in the vortex
  core (the apparent nonzero minimum is due to finite numerical
  resolution) which is filled by $\zeta_0^{(x)}$, keeping the density
  nonzero everywhere.}
  \label{fig:split-core-1D}
\end{figure}

After demonstrating that the singular singly quantized vortex of
Eq.~(\ref{eq:fm-singular}) is energetically
stable, we next study its vortex core structure after the energy
relaxation. The resulting vortex configuration with a stable vortex
core is shown in Fig.~\ref{fig:split-core-3D}(a). The vortex lines in the
different spinor wavefunction components have moved apart and no
longer spatially overlap. We show in Fig.~\ref{fig:split-core-1D}(a)
a 1D density cut along which the spatially separated vortices are
aligned. The vortex line of the $\zeta_0$ component is located at the
center of
the trap and the vortices of the $\zeta_{\pm}$ components are
symmetrically displaced from the center. This split-core solution
appears to break the explicit axial symmetry of the spinor
  component densities in
Eq.~(\ref{eq:fm-singular}). A similar core splitting has previously
been demonstrated in
2D numerical simulations in Ref.~\cite{mizushima_pra_2002}.
We will show below how it is beneficial to analyze the vortex core
using a spinor basis transformation. In particular, after an appropriate
transformation we can easily identify the location of the vortex, the nonvanishing
atom density at the vortex line singularity, and the axially symmetric density
profiles of the spinor components in the new basis representation.
In the vortex configuration displayed in Fig.~\ref{fig:split-core-3D}(a) we may then
identify the split-core vortex as spin winding around a core of nonvanishing atom
density.

\begin{figure}[tb]
  \centering
  \includegraphics[scale=1]{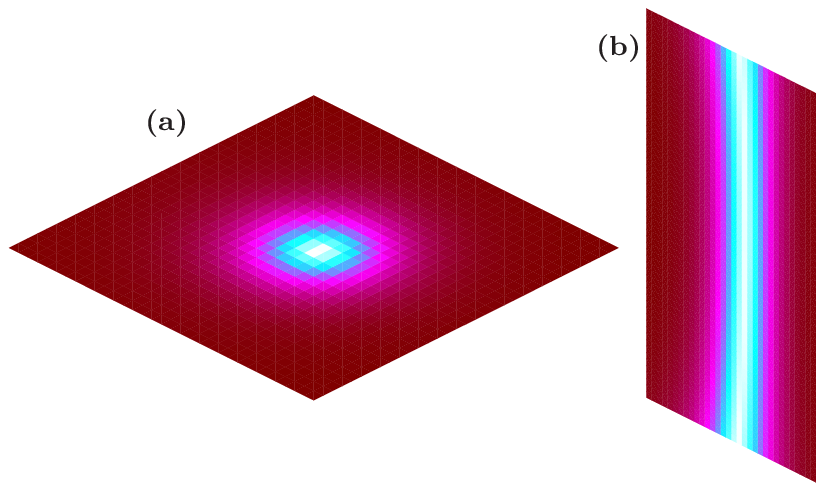}
  \caption{(color online) Spin magnitude $\absF$, showing the core of
    the stable
    singular vortex of Fig.~\ref{fig:split-core-1D} in the FM phase,
    shown in the $xy$ plane (a) and
    $yz$ plane (b). Outside the vortex core, $\absF=1$ (dark red/dark gray) in
    the FM order parameter manifold.  In the core, the singularity is
    accommodated by enforcing $\absF=0$ (white) while maintaining
    nonzero density.  The size of the
    core region is determined by the spin healing length
    $\xi_F$.}
    \label{fig:fm-vortex-core}
\end{figure}

In order to analyze the vortex configuration of
Fig.~\ref{fig:split-core-3D}(a) we perform a basis transformation for the
spinor wavefunction. We transform the split-core spinor to the
basis where spin is quantized along the $x$ axis as
$\zeta^{(x)}=U^{-1}(0,\pi/2,0)\zeta^{(z)}$, explicitly indicating
the spinor basis by superscripts.  In $\zeta^{(x)}$ the vortex appears
as an
opposite winding of the phase in the two components
$\zeta^{(x)}_\pm$. These vortex lines
again overlap as shown in Figs.~\ref{fig:split-core-3D}(b) and
\ref{fig:split-core-1D}(b).
Crucially, there is no vortex line in $\zeta^{(x)}_0$, and this component
therefore fills the vortex cores of the two other components so that
the density is nonvanishing everywhere. The single vortex core,
which is readily apparent from Fig.~\ref{fig:fm-vortex-core}, is
thus explicitly restored in $\zeta^{(x)}$ by the transformation to the
`natural basis' of the vortex.  We identify the spinor wavefunction
resulting from the basis transformation now as having the same
structure as the singular vortex,
defined in Eq.~(\ref{eq:spinvortex}).  In Appendix~\ref{append} we
show how the core deformation and the relation between $\zeta^{(x)}$
and $\zeta^{(z)}$ can be understood qualitatively through an analytic
treatment.

\begin{figure}[tb]
  \centering
  \includegraphics[scale=1]{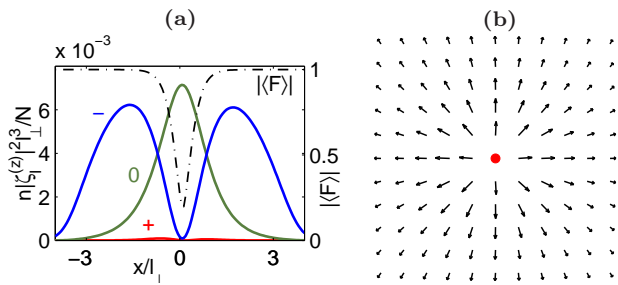}
\caption{(color online) Axially symmetric spin vortex. (a)
  Densities of the spinor components together with the spin magnitude
  along a radial cut (lines and labels as in
  Fig.~\ref{fig:split-core-1D}(a)). The vortex lines in $\zeta_\pm$
  overlap
  perfectly at the position of the vortex core. (b) Relaxed spin
  profile in the $xy$ plane, showing the characteristic radial
  disgyration of the spin
  vector around the singular core. At large radii the spin vector
  bends out of the $xy$ plane. The vortex line singularity is marked
  by a dot at the center.}
  \label{fig:spinvortex}
\end{figure}

The spin structure of the
stable vortex is shown in Fig.~\ref{fig:cl-v-singular}(b).
The vortex line is oriented along the $z$ direction---the axis about which
the trap is rotating. However, the spinor takes the form of
Eq.~(\ref{eq:spinvortex}) in the basis defined along the (co-rotating) $x$ axis.
This vortex is singular, preserving the topology, and can be
reached from Eq.~(\ref{eq:fm-singular}) by local spin rotations and
could similarly be continuously transformed into the singular spin vortex
[Fig.~\ref{fig:fm-vortex-schematic}(b)]. The stable vortex core Fig.~\ref{fig:cl-v-singular}(b)
has a broken spatial parity (the spin profile has an antisymmetric spatial parity close to
the vortex core). This spin profile is nonaxisymmetric. We also find a stable axisymmetric vortex core.
This is achieved by starting energy relaxation from
Eq.~(\ref{eq:spinvortex}), such that the radial disgyration of the
spin vector is present already in the initial state.  The spinor
components and the resulting spin profile are shown in
Fig.~\ref{fig:spinvortex}. The dependence of the final configuration on the initial state indicates
a close energetic degeneracy of the two solutions.

To understand the vortex core deformation it is also beneficial to compare the
initial-state singular vortex of Eq.~(\ref{eq:fm-singular}) to the
vortex obtained in the energy minimization. In
Eq.~(\ref{eq:fm-singular}) each spinor component exhibits a singly
quantized vortex.
These overlapping vortex lines imply that the total density $n(\mathbf{r})$
must be zero on the singular line in order to maintain
single-valuedness of the order parameter.
The size of the vortex core is determined by the healing length $\xi_n$.
The density depletion can be avoided by splitting the vortex
core such that the vortex lines in the spinor components no longer
overlap.
Since the total condensate density then does not vanish at the vortex line
where the order parameter is singular, we must now require that the spinor wavefunction
becomes orthogonal to the ground-state manifold at the vortex singularity.
In the FM manifold $\absF=1$, so at the vortex line we must have
$\absF=0$, which represents the spin magnitude of the polar phase.

The spin magnitude of the numerically calculated singular vortex core
is displayed in Figs.~\ref{fig:split-core-1D}(b) and \ref{fig:fm-vortex-core}.
We find that the value of the spin magnitude indeed rapidly approaches
zero close to the vortex line singularity
(the small deviations from zero are due to spatial resolution of the numerics).
This indicates the formation of a polar vortex core, constituting a local violation of the spin condition
for the ground-state FM manifold. An analytic description of the vortex solution is provided in Appendix~\ref{append}.
The size of the vortex core is determined by the spin healing
length $\xi_F$.  The splitting is then energetically favorable when
$\xi_F$ allows a larger core size (i.e., when $\xi_F\agt \xi_n$) such that the energy cost of
violating $\absF=1$ is smaller than that of depleting the density.

We find that the region where the spin magnitude deviates from
$\absF=1$ extends over the entire core size,
determined by the spin healing length $\xi_F$. Outside the core region
of the vortex the symmetry of the spin-1 BEC is broken according to
the FM energy condition of the spin-dependent interaction energy, so
that we have $\absF=1$. Close to the singular vortex, however, the
order-parameter bending energy restores the symmetry of the full
spin-1 BEC wavefunction ($S^5$ determined by a normalized spinor
wavefunction of three complex components), mixing the FM and polar phases. The bending
energy is enhanced very close to the vortex singularity due to the
large density gradient contributions that excite the system from the
FM ground-state manifold. An analogous core deformation was previously
found for a singular point defect in a polar spin-1 BEC in
Ref.~\cite{ruostekoski_monopole_2003}. In that case an isotropic point
defect with a vanishing density deformed to a ring defect with a FM
core.  This effect is closely related to the deformation of the core
of a singular vortex in the polar phase described in Sec.~\ref{sec:polar-vortex}.

In experiments a stable singular vortex could be prepared in a
controlled way by phase-imprinting the initial
singular vortex state of Eq.~(\ref{eq:fm-singular}) in a rotating
trap, so that the parameter values of the system belong to the stable
region of the stability diagram displayed in
Fig.~\ref{fig:isotropic-stability}(b). The initial-state vortex
[Eq.~(\ref{eq:fm-singular})] is composed by perfectly overlapping
singly quantized vortices in each of the spinor components. These could
be phase-imprinted using previously realized experimental
techniques~\cite{matthews_prl_1999,leanhardt02,andersen_prl_2006}.
The stability diagram also indicates the
conditions under which a singular vortex created in a phase
transition~\cite{sadler_nature_2006} could potentially be
stabilized.

In the above analysis, we have allowed the magnetization $M_z=N_+-N_-$,
where $N_\pm$ are the total populations of $\zeta_\pm$, to vary during
the relaxation process.
This in principle allows a
spontaneous magnetization to develop in the system.
In experiments, dissipative relaxation of energy due to atomic
collisions may frequently conserve the magnetization.  We have
therefore also performed calculations where a weak magnetization is enforced
throughout the energy-minimization procedure. We find the relaxed vortex configurations
of Figs.~\ref{fig:split-core-3D}-\ref{fig:fm-vortex-core}, however,
qualitatively similar to the ones where no fixed magnetization was forced.

\begin{figure}
 \centering
 \includegraphics[scale=1]{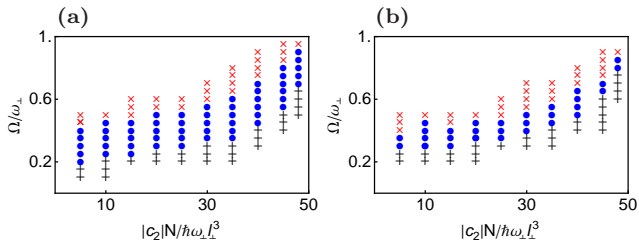}
 \caption{(color online) Stability of the FM coreless (a) and singular
   (b) vortices
   in a highly oblate trap, $\omega_{\perp}/\omega_z=0.1$,
   with $Nc_0=50\hbar\omega_{\perp}l_{\perp}^3$ and
   $-48\hbar\omega_{\perp}l_{\perp}^3 \leq Nc_2 \leq
   -5\hbar\omega_{\perp}l_{\perp}^3$.
   (Symbols as in Fig.~\ref{fig:isotropic-stability}.)}
 \label{fig:pancake-stability}
\end{figure}

Thus far we have considered an isotropic trap.  We find that the results are
qualitatively similar in an oblate trap with
$\omega_\perp/\omega_z=0.1$. We find that also in this regime, the singular vortex represents a local
energetic minimum and is stable for a range of $\Omega$, again despite
the fact that a lower-energy coreless-vortex solution exists.  The
parameter regions allowing stable nonsingular coreless and singular
split-core vortices in the oblate trap are shown in
Fig.~\ref{fig:pancake-stability}.
In the oblate trap case the simulations with fixed magnetization produce
more symmetric vortex configurations than in the simulations where the magnetization varies freely
but the qualitative features are the same in the two cases.

\begin{figure}
  \centering
  \includegraphics[scale=1]{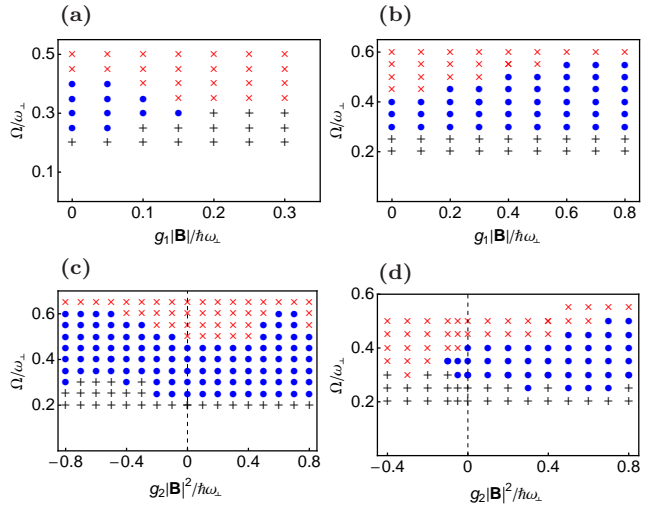}
  \caption{(color online) Effects of linear and quadratic Zeeman
    splitting on the
    stability of the FM vortices in an oblate trap
    ($\omega_\perp/\omega_z=0.1$).  In all panels
    $Nc_0=50\hbar\omega_{\perp}l_{\perp}^3$ and
    $Nc_2=-10\hbar\omega_{\perp}l_{\perp}^3$.
    (a) Stability of a coreless vortex in the presence of
    linear Zeeman splitting. The vortex state becomes unstable for
    $g_1\abs{\mathbf{B}} \agt 0.2\hbar\omega_\perp$.
    (b) The singular vortex remains stable despite linear Zeeman splitting.
    (c) Stability of the coreless vortex in the presence of quadratic
    Zeeman splitting.  A stable region is found at all investigated
    values of $g_2\abs{\mathbf{B}}$.
    (d) Stability of the singular vortex in the presence of quadratic Zeeman
    splitting. The vortex becomes unstable for a relatively small negative
    $g_2\abs{\mathbf{B}}$.
    (Symbols as in Fig.~\ref{fig:isotropic-stability}.)}
  \label{fig:zeeman}
\end{figure}

Applying a weak external magnetic field introduces a Zeeman shift
between the spinor components according to
Eq.~(\ref{eq:hamiltonian-density}).  The spinor nature of the BEC is
retained as long as the applied field is not too strong $g_1\!\abs{\mathbf{B}},g_2\!\abs{\mathbf{B}}^2\alt \mu$, where $\mu$ denotes the chemical potential.
In the case of a small linear Zeeman splitting
$g_1\!\abs{\mathbf{B}}$ (taking $\mathbf{B}$ along the $z$ axis) we find that
the coreless and singular vortices are both stable, with the coreless vortex lower in
energy.  The Zeeman splitting will tend to align the spins with the
applied field.  This causes the energy
of the coreless vortex to increase as maintaining the fountain-like
spin structure becomes energetically  less favorable.  Thus we find that for
$g_1\!\abs{\mathbf{B}}\gtrsim0.2\hbar\omega_{\perp}$ the coreless vortex is no
longer stable.  The
singular vortex, on the other hand, remains energetically stable for
all $g_1\!\abs{\mathbf{B}}$ considered (up to $0.8\hbar\omega_{\perp}$), as shown in
Figs.~\ref{fig:zeeman}(a) and (b). For a sufficiently large linear Zeeman splitting
the ideal spinor basis to analyze the singular vortex core becomes the one defined by the magnetic field.

A quadratic Zeeman splitting, on the other hand, does not destroy the
stability of the coreless vortex, but for
$g_2\!\abs{\mathbf{B}}^2 \alt -0.1\hbar\omega_{\perp}$ the
singular vortex is no longer energetically stable
[Fig.~\ref{fig:zeeman}(c) and (d)]. For a sufficiently large positive quadratic Zeeman
splitting the ideal spinor basis to analyze the singular vortex core is oriented perpendicular to the magnetic field.

\section{Stability and core structure of a polar vortex}
\label{sec:polar-vortex}

We now turn our attention to the polar phase of a spin-1 BEC. In this case the interaction term
$c_2>0$ in the Hamiltonian \eqref{eq:hamiltonian-density}, and it is energetically favorable to minimize
the spin magnitude everywhere in space so that $\absF=0$. We take a
representative spinor $\zeta = (0,1,0)^T$ whose macroscopic condensate
spin quantization axis is oriented along the $z$ axis. The general
spinor wavefunction may then be constructed from the macroscopic
condensate phase $\phi$ and the spin rotations defined by the Euler angles
$(\alpha,\beta,\gamma)$ as
\begin{equation}
  \label{eq:polar}
  \zeta^{\rm p} =
  e^{i\phi}U(\alpha,\beta,\gamma)\threevec{0}{1}{0}
  = \frac{e^{i\phi}}{\sqrt{2}}\threevec{-e^{-i\alpha}\sin\beta}
                     {\sqrt{2}\cos\beta}
                     {e^{i\alpha}\sin\beta}\,.
\end{equation}
It is beneficial to introduce the unit vector $\nematic = (\cos\alpha\sin\beta,\sin\alpha\sin\beta,\cos\beta)$ that defines the local direction of the condensate spin quantization. We may then write  the spinor wavefunction in terms of $\nematic$ as~\cite{ruostekoski_monopole_2003}
\begin{equation}
  \zeta^\mathrm{p} = \frac{e^{i\phi}}{\sqrt{2}}
                    \threevec{-d_x+id_y}{\sqrt{2}d_z}{d_x+id_y}\,.
\end{equation}
The unit vector $\nematic$ takes values on a sphere and the condensate phase $\phi$ on a unit circle. The state of the spinor wavefunction, however, remains unchanged when a $\pi$-rotation of $\phi$ is combined with inverting the $\nematic$ vector, so that the states
$\zeta^{\rm p}(\phi,\nematic) = \zeta^{\rm p}(\phi+\pi,-\nematic)$ are identical.
These states must be identified to avoid double counting, and the
order-parameter space
is therefore $(\U(1) \times S^2)/\mathbb{Z}_2$, from the condensate
phase and rotations of $\nematic$, factorized by the
discrete two-element group $\mathbb{Z}_2$ due to the identification.
The vector $\nematic$ is thus taken to be unoriented and defines a
nematic axis~\cite{zhou_ijmpb_2003}.

A singly quantized vortex in the polar phase can be formed as a $2\pi$-winding of
the condensate phase $\phi$ around a closed loop encircling the vortex core. Choosing the vortex line
along the $z$ axis, we obtain
\begin{equation}
  \label{eq:p-vortex}
  \zeta^\mathrm{1} =
  \frac{e^{i\varphi}}{\sqrt{2}}\threevec{-e^{-i\alpha}\sin\beta}
                     {\sqrt{2}\cos\beta}
                     {e^{i\alpha}\sin\beta}\,.
\end{equation}
However, the nematic order also allows the formation of a vortex
carrying half a quantum of circulation~\cite{leonhardt_jetplett_2000} (in an analogy to half-quantum vortices in superfluid liquid $^3$He~\cite{vollhardt-wolfle}), constructed as a $\pi$-winding of the
macroscopic condensate phase together with a $\nematic\to-\nematic$
rotation of the nematic axis around a closed loop encircling the vortex core. For example, we may have
\begin{equation}
  \label{eq:hq}
  \zeta^\mathrm{1/2} = \frac{e^{i\varphi/2}}{\sqrt{2}}
                   \threevec{-e^{-i\varphi/2}}
                            {0}
                            {e^{i\varphi/2}}
		 = \frac{1}{\sqrt{2}}
                   \threevec{-1}
                            {0}
                            {e^{i\varphi}}\,.
\end{equation}
Circulation is thus quantized in units of $\pi$, half the circulation
of Eq.~(\ref{eq:p-vortex}). This will be a crucial observation when we now analyze
possible deformations of the core of a singly quantized vortex as energy is
minimized.

\begin{figure}
  \includegraphics[scale=1]{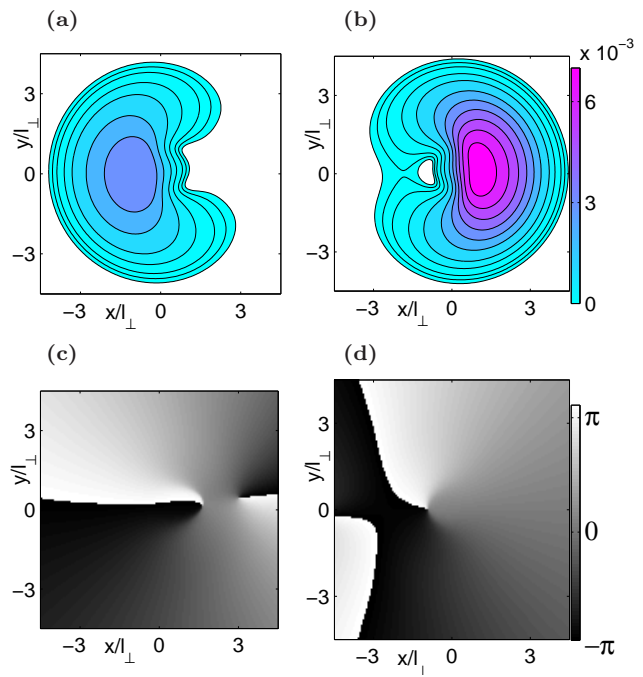}
  \caption{(color online) Stable core structure of the singular vortex
  in the polar
  phase shown in the $xy$ plane. (a) and (b): Densities in
  $\zeta_+^{(z)}$ and $\zeta_0^{(z)}$, respectively. (c) and (d): The
  corresponding phases.
  ($\zeta_-^{(z)}$ is identical to $\zeta_+^{(z)}$
  up to a global $\pi$ phase shift.)
  The spinor wavefunction exhibits vortex lines with highly deformed
  anisotropic cores in the spinor components.}
  \label{fig:polar-split-core}
\end{figure}

In order to investigate the energetic stability of a singly quantized
singular vortex in the polar phase
of a spin-1 BEC we numerically minimize the energy of the system in a rotating frame. We follow the same
procedure as in the FM case and this time take a singular polar vortex
of Eq.~(\ref{eq:p-vortex}) with $\beta=\pi/4$ and $\alpha=0$ as an
initial state of the numerical relaxation. Similarly as in the FM case
of Eq~(\ref{eq:fm-singular}), the initial state is formed by
overlapping vortex lines in all the three spinor components. Upon
minimizing the energy, the vortex cores of the individual spinor
components separate.
However, compared with the FM case, the splitting is now more
complicated, as shown in Fig.~\ref{fig:polar-split-core}. The result
is highly deformed anisotropic vortex cores in the spinor
components. The vortices in $\zeta_+$ and
$\zeta_-$ overlap, but the one in $\zeta_0$ is displaced from the
other two. There are no simultaneous density
minima in all three spinor components, and the density is therefore nonzero
everywhere.
Similar split-core solutions found by numerical calculation in a rotating
2D system~\cite{yip_prl_1999} have resulted in some controversy regarding the number vortices
in the individual spinor components in the final
configuration~\cite{mizushima_pra_2002}.  In the previous
2D studies the stable core structures were not classified. Here we
show now how the split core in
Fig.~\ref{fig:polar-split-core} can be identified as a
topology-preserving splitting of the singly quantized vortex into a
pair of half-quantum vortices as illustrated schematically in
Fig.~\ref{fig:core-splitting-schematic}.

\begin{figure}
  \includegraphics[width=0.99\columnwidth]{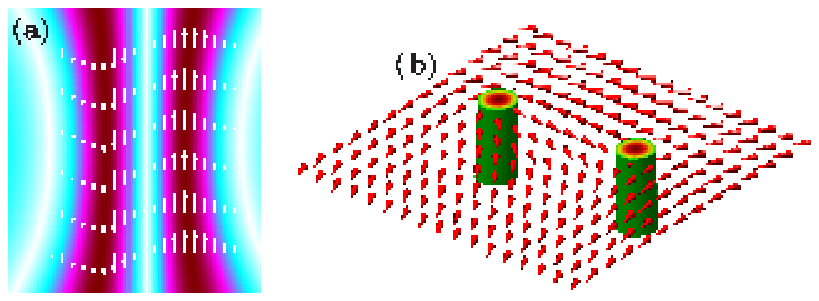}
  \caption{(color online) Splitting of the singly quantized vortex
  into two half-quantum vortices.  (a) Spin magnitude $\absF$ [color
  map from white ($\absF=0$) to red/dark gray ($\absF=1$)] together
  with the spin vector (arrows), showing the FM cores
  with nonvanishing density. The spins are antiparallel in the
  two cores.
  (b) Nematic axis $\nematic ({\rm r})$ together with the
  vortex cores (indicated by green/light gray isosurfaces of $\absF$, with
  increasing spin magnitude indicated by the color gradient inside).  Away
  from the vortex cores the topology of the initial singly quantized
  vortex is preserved.  In the core region, $\nematic$ winds
  by $\pi$ about each half-quantum vortex core. For visualization
  purposes, the unoriented $\nematic$-field is shown as cones.
  Here a quadratic Zeeman shift has been
  introduced to ensure that $\nematic$ lies in the $xy$ plane and the
  spins align with the $z$ axis.}
  \label{fig:fm-cores}
\end{figure}

In the numerical simulations the initial state of a singly quantized
singular vortex in Eq.~(\ref{eq:p-vortex}) is composed of three
perfectly overlapping vortex lines in each of the three spinor
components. The polar vortex consequently has a vanishing density at
the line singularity of the polar order parameter of the spin-1
BEC. The singular vortex with zero density is energetically unstable
with respect to core deformation.
As the vortices of the individual spinor components move apart during
energy relaxation, the density becomes nonvanishing everywhere in the
vortex-core region. Similarly to the FM vortex case, we must therefore
require that the spinor wavefunction
becomes orthogonal to the ground-state manifold at the vortex
singularity. This indicates that we must have $\absF=1$ on the vortex
line. We show in Fig.~\ref{fig:fm-cores} the numerically calculated
vortex core structure of a stable vortex whose initial state is the
singular singly quantized vortex of Eq.~(\ref{eq:p-vortex}). The
displayed spin magnitude exhibits two clearly separated cores in which
the peak value increases to $\absF=1$, indicating the emergence of a
FM core region for the vortex.

The formation of the FM cores can be understood from the same argument used to understand the
polar core of the singular FM vortex in Sec.~\ref{sec:fm-vortex} and is illustrated in Fig.~\ref{fig:core-splitting-schematic}:
The singular polar vortex (\ref{eq:p-vortex}), which is used as an initial state in the energy relaxation,
implies a density-depleted core whose size
is determined by $\xi_n$. However, accommodating a singularity of the polar order parameter by requiring $\absF=1$ at the vortex line
means that the length scale, and thus the associated bending energy,
is determined by $\xi_F$. The energy of Eq.~(\ref{eq:p-vortex}) can
thus be lowered by having a nonvanishing atom density and by extending the core size from $\xi_n$ to $\xi_F$.
In the case of a polar vortex this is achieved by spontaneously breaking the axial symmetry and
forming two FM cores by the mechanism sketched in
Fig.~\ref{fig:core-splitting-schematic}. The separation between the
cores is of the order of $\xi_F$, depending also on the angular
momentum of the system when it adjusts to the rotation frequency and
on the density gradient due to the trap.

\begin{figure}
  \includegraphics[scale=1]{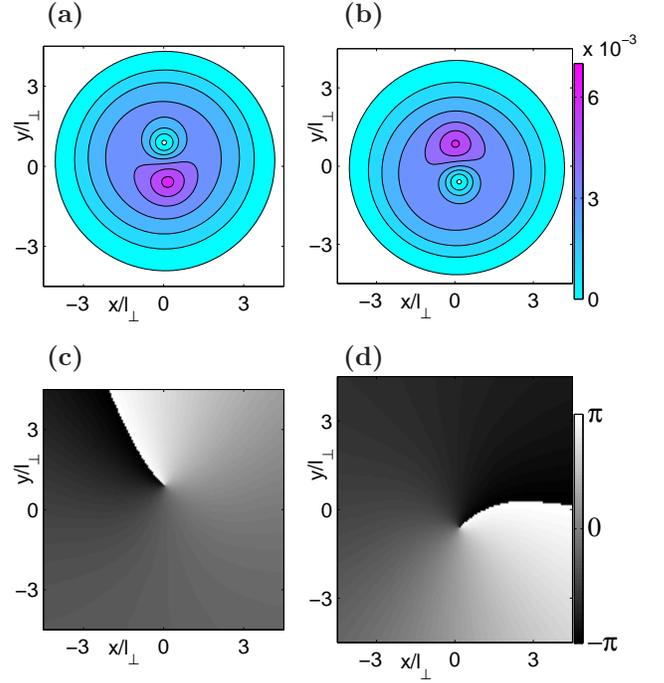}
  \caption{(color online) Spinor wavefunction of the stable singular
    vortex state
    from Fig.~\ref{fig:polar-split-core} after spinor basis
    transformation such that spin is quantized along the $y$ axis.
    (a) and (b): Densities in $\zeta_+^{(y)}$ and $\zeta_-^{(y)}$,
    respectively. (c) and (d): The corresponding phases. The component
    $\zeta_0^{(y)}$ (not shown) is unpopulated.
    The previously complex structure can now be identified as a pair
    of half-quantum vortices.}
  \label{fig:polar-rotated}
\end{figure}

\begin{figure}
  \includegraphics[scale=1]{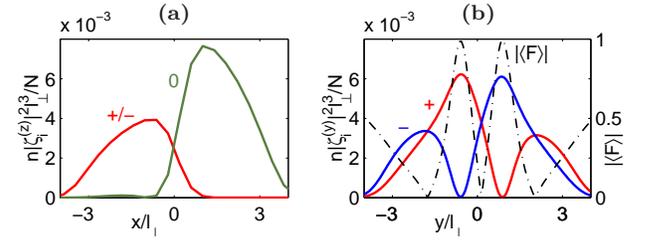}
  \caption{(color online) (a) Density profiles of the spinor
    components on the axis
    connecting their density minima
    (cf.\ Fig.~\ref{fig:polar-split-core}). Lines labeled with spinor
    component index. Note that $|\zeta_\pm|$ exactly overlap.
    (b) Spinor-component density
    profiles in $\zeta^{(y)}$ after basis transformation (cf.\
    Fig.~\ref{fig:polar-split-core}) plotted along the axis connecting
    the half-quantum vortices. The spin magnitude $\absF$ (black
    dash-dotted line) shows the FM cores. Unpopulated $\zeta^{(y)}_0$
    is not shown.}
  \label{fig:polar-1D}
\end{figure}

We may analyze this symmetry breaking of the vortex core by means of a basis
transformation. We write the spinor in the basis of spin projection
onto the axis given by the spin vector in the FM core.
For the case of Fig.~\ref{fig:polar-split-core}(a),
the spins in the two cores align (antialign) with the $y$ axis and we calculate
$\zeta^{(y)}=U^{-1}(\pi/2,\pi/2,0)\zeta^{(z)}$.  The resulting spinor then
shows displaced vortex lines in $\zeta^{(y)}_\pm$ while the density
vanishes in $\zeta^{(y)}_0$ [Fig.~\ref{fig:polar-rotated}].  In the
new spinor basis, the vortex lines in
$\zeta^{(y)}_\pm$ coincide precisely with the spin maxima, as shown in
Fig.~\ref{fig:polar-1D}

We can now identify the core structure emerging from the splitting of the singular vortex by comparing
the spin-rotated state $\zeta^{(y)}$ with Eq.~(\ref{eq:hq}). We then find that each vortex line in
$\zeta^{(y)}$ has exactly the form of a half-quantum vortex. The split-core configuration may thus be
interpreted as a splitting of the singly quantized vortex into a pair of half-quantum vortices with
FM cores. The topological charges of vortices are here additive
and topology is therefore preserved when the
singly quantized vortex splits into the pair of half-quantum vortices.
This can also be inferred from the behavior of the nematic axis
$\nematic$.  Figure~\ref{fig:fm-cores}(b) shows $\nematic$ in a
numerical solution together with the
FM cores of the half-quantum vortices.  Away from the
vortices, there is no net winding in $\nematic$ on a path enclosing the
vortices.  However on a path that encircles only one vortex core,
$\nematic$ turns by $\pi$, indicating the emergence of a disclination
plane as indicated in Fig.~\ref{fig:core-splitting-schematic}.

As in the case of a FM vortex, the core deformation can be explained
in terms of the vortex topology and the energetic hierarchy of
different length scales (see Fig.~\ref{fig:core-splitting-schematic}).
Outside the vortex core region of size $\xi_F$, where the order
parameter bending energy is not sufficient to excite the system away
from the polar ground-state manifold, we have $\absF=0$ and the
topological properties of the initial singly quantized singular vortex
are preserved. This is indicated by the unit winding of the
macroscopic condensate phase around any closed loop encircling the
entire vortex core and by the nematic vector field outside the core
region. It is only inside the core of size $\xi_F$ that the strong
order parameter bending energy restores the symmetry of the full
spin-1 condensate wavefunction by exciting the system out of the polar
ground-state manifold and by allowing the complete range of spin
values $\absF$ from 0 to 1. The local deformation of the core is
topologically possible due to the nematic order of the polar phase, where
the axis $\nematic$ is unoriented, with the opposite orientations
$\nematic=-\nematic$ identified. The core deformation mechanism of the
vortex line is related to the point defect deformation into a singular
ring where the nematic order allows the spontaneous breaking of the
spherical defect core symmetry~\cite{ruostekoski_monopole_2003}.
In the $B$-phase of superfluid liquid $^3$He, a stable nonaxisymmetric
singular vortex with
a nonzero superfluid density at the core was theoretically predicted in Refs.~\cite{thuneberg_prl_1986,salomaa_prl_1986} and
experimentally observed in Ref.~\cite{kondo_prl_1991}. The $^3$He $A$-phase core
was explained to consist of two half-quantum vortices. In the high-pressure regime
the axial symmetry of the vortex is restored but the core can still remain in the $A$-phase
with a nonvanishing superfluid density~\cite{salomaa_prl_1983}.

\begin{figure}
 \centering
 \includegraphics[scale=1]{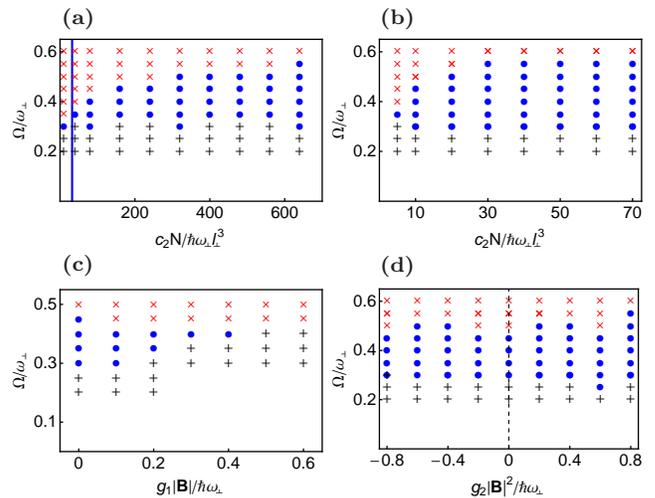}
 \caption{(color online) Energetic stability of the split-core singular
   vortex in the polar phase.
   (a) Stability in the isotropic trap for varying $c_2$ using
   $Nc_0=1000\hbar\omega_{\perp}l_{\perp}^3$.  The vertical line
   marks the value
   $c_0/c_2\simeq 31$ for $^{23}$Na~\cite{crubellier_epjd_1999}.
   (b) Stability in the oblate trap ($\omega_{\perp}/\omega_z=0.1$) for
   varying $c_2$ using $Nc_0=50\hbar\omega_{\perp}l_{\perp}^3$.
   (c) and (d): Stability of the split-core singular vortex in the
   presence of linear and quadratic Zeeman splitting, respectively.
   Increasing linear Zeeman splitting renders the singular vortex
   unstable, whereas the stability is robust against quadratic Zeeman
   splitting. In the slowly rotating region for all panels, the
   instability of the split-core singular vortex may be either towards
   the vortex-free state, or towards the state with a single
   half-quantum vortex. (Symbols as in Fig.~\ref{fig:isotropic-stability}.)}
 \label{fig:polar-stability}
\end{figure}

We find that the singular vortex splits into a pair of half-quantum
vortices by the mechanism described above for all investigated
parameter regimes.  However, we find a critical rotation frequency of
$\Omega \simeq 0.3\omega_\perp$ below which the vortices
start exiting the atom cloud.
Figure~\ref{fig:polar-stability} shows the energetic stability of the
half-quantum vortex pair obtained from splitting of a singly quantized
vortex in both isotropic and oblate ($\omega_{\perp}/\omega_z=0.1$) traps.
The energetic ground state of a rotating polar spin-1 BEC can consist of
half-quantum vortices~\cite{reijnders_pra_2004,mueller_pra_2004}, and
increasing the rotation frequency leads to nucleation
of more half-quantum vortices in addition to the split core of the
initial singular vortex.

The splitting mechanism of the singly quantized vortex is qualitatively similar also when a weak
Zeeman splitting due to a magnetic field is introduced.  However, as shown in
Fig.~\ref{fig:polar-stability}(c), a linear Zeeman splitting of
$g_1B\gtrsim 0.4\hbar\omega_\perp$ causes the resulting pair of
half-quantum vortices to become energetically unstable at all
rotational frequencies.  By contrast, the vortex pair remains stable
above $\Omega \simeq 0.3 \omega_\perp$ for the entire range of
quadratic Zeeman splittings considered
($-0.8\hbar\omega_{\perp} \leq g_2|B|^2\leq0.8\hbar\omega_{\perp}$)
 [Fig.~\ref{fig:polar-stability}(d)].

\section{Conclusions}
\label{sec:conclusions}

We have demonstrated that a singular, singly quantized vortex can be
energetically stable in the FM phase of a spin-1 BEC, despite
the existence of a stable coreless vortex with lower energy (increasing the rotation frequency in a vortex-free BEC also leads to nucleation of coreless vortices).
This implies that even though singular vortices would not be
nucleated by rotation alone, a singly quantized vortex created, for
example, by phase-imprinting would remain stable in the rotating system.
This provides an interesting opportunity for controlled studies of a singular
vortex line in a ground-state manifold with a broken $\SO(3)$ symmetry.
Such a system supports only two topological classes of vortices: those that can be locally
deformed to a vortex-free configuration and those that are topologically equivalent
to a singly quantized singular vortex. Experimentally, one could phase-imprint
overlapping vortex lines in each of the three spinor wavefunction components.
The resulting structure represents a singular spin-1 vortex with a vanishing density at the vortex line.
The core of such a vortex then deforms to a energy-minimized configuration within
the same topological equivalence class.

The stable vortex core in the FM phase is formed by nonoverlapping vortex lines in the
three spinor components.  We have demonstrated that this
seemingly complex core structure can be understood in terms of
the combination of the vortex topology and the
energetics of characteristic
length scales. By deforming the core of a
singly quantized, singular vortex in the FM phase so as to maintain a
nonzero density everywhere, instead accommodating the singularity by
forcing $\absF=0$, the gradient contribution to the energy is
lowered.  The reason is that the size of the defect core is then
determined by the spin healing length $\xi_F$ which is in general
larger than the characteristic size $\xi_n$ of a defect core where the
density goes to zero.
In other words, in the larger core size case with a nonvanishing atom density, the gradient energy restores the full symmetry of the spin-1 condensate wavefunction within the core region. The system then simultaneously exhibits two different order parameter symmetries: maximal unbroken symmetry inside the core of size $\xi_F$ and a broken symmetry (of the FM phase) outside the vortex core.

The core deformation mechanism results in a singular vortex whose
core is also filled with atoms in the polar phase.  The spin vector winds by $2\pi$
as the core is encircled.  The single vortex core
can be explicitly restored in the spinor by judicious choice of spinor
basis.

In the polar phase, we have shown that a singly quantized vortex is
stabilized by a spontaneous breaking of axial symmetry.  The resulting
stable defect is a pair of half-quantum vortices with FM cores, which
is stable in a sufficiently rapidly rotating trap.  The
formation of the FM cores avoids depleting the density in the vortex
core. This is energetically favorable by the same reasoning that was
applied to explain the polar core of the singular vortex in the FM
phase. The resulting spinor wavefunction is
analyzed and the vortex structure identified through a rotation of the
spinor basis, so that in the rotated basis the half-quantum
vortices appear as separate vortex lines in the $\zeta_\pm$
components.

\begin{acknowledgments}
  We acknowledge financial support from the EPSRC and the Leverhulme Trust.
\end{acknowledgments}

\appendix
\section{Basis transformation for FM vortex}
\label{append}

In the numerical simulations we found that the singular FM vortex relaxes
to a stable configuration formed by nonoverlapping vortex lines in the three
spinor components, as shown in Fig.~\ref{fig:split-core-3D}(a).
By an appropriately chosen basis transformation we showed that
this seemingly complex vortex structure can be identified as a single, singular vortex
with the line singularity populated by atoms in the polar phase [Fig.~\ref{fig:split-core-3D}(b)].
In this Appendix we demonstrate this basis transformation
through a qualitative analytic treatment and show how the core structure of the
singular FM vortex may be identified.

For simplicity, we implement the basis transformation by starting from the final
configuration of the singular vortex with a single vortex line as in Figs.~\ref{fig:split-core-3D}(b)
and~\ref{fig:spinvortex} and rotating to the configuration of nonoverlapping vortex lines in the three
spinor components [see, e.g., Fig.~\ref{fig:split-core-3D}(a)]. The analytic expressions become notably simpler in the case of the axisymmetric vortex of Fig.~\ref{fig:spinvortex} than with the one displaying a more complex spin rotation in Figs.~\ref{fig:}(b) and \ref{fig:split-core-3D}, but the basic principle of the transformation is the same in both cases. In order to describe the vortex of Fig.~\ref{fig:spinvortex} we rewrite the singular vortex
displaying a radial disgyration of the spin vector $\zeta^{\rm sv}$ of Eq.~(\ref{eq:spinvortex})
in the following form
\begin{equation}
  \zeta =
  \frac{1}{\sqrt{2}}\threevec{\sqrt{2}(\cos\varphi-i\sin\varphi)f\cos^2\frac{\beta}{2}}
                             {g \sin\beta}
                             {\sqrt{2}(\cos\varphi+i\sin\varphi)f\sin^2\frac{\beta}{2}}\,.
                             \label{eq:parametereq}
\end{equation}
Here we have introduced the profile function $f(x,y)$ and the notation
\begin{equation}
g \sin\beta =\sqrt{2-f^2\left(1+\cos^2\beta\right)}\,.
\end{equation}
The profile function $f$ describes the mixing of the FM and the polar phases in the core region
of the vortex. For $\beta\neq \pi$ we obtain a radial disgyration with nonvanishing density
at the singularity (for $\beta=\pi$ and $f=1$ we recover a vortex with zero density at the singularity). We assume $f$ to be monotonically increasing from 0 at the $z$ axis (the vortex singularity) reaching 1 outside the vortex core of size $\xi_F$. For the numerically minimized stable solution of Fig.~\ref{fig:spinvortex} the parameter $\beta$ is not constant; close to the vortex core we have $\beta\simeq \pi/2$ and far away from the vortex $\beta\rightarrow\pi$.
The magnitude of the spin vector may be evaluated from Eq.~(\ref{eq:parametereq}), yielding $\absF=f\sqrt{2-f^2}$. At the line singularity $\absF=0$ representing the polar phase and outside the core region $\absF=1$ that corresponds to the FM phase. In between $\absF$ continuously varies between these two values, indicating the mixing of the two phases.

When we perform the rotation of Eq.~(\ref{eq:parametereq}) by the angle of $-\pi/2$ with respect to the $x$ axis, we obtain the spinor wavefunction
\begin{equation}
  \zeta^{(-x)} =\frac{1}{2}\threevec{f \left(\cos\varphi-i\sin\varphi\cos\beta \right) + g \sin\beta}
{\sqrt{2}f \left(-\cos\varphi\cos\beta + i\sin\varphi \right)}
{f \left(\cos\varphi - i\sin\varphi\cos\beta \right) - g \sin\beta}\,.\label{eq;parameterfinal}
\end{equation}
The spinor wavefunctions are of the form $(x-x_0) +i \eta (y-y_0)$, indicating that a singly
quantized vortex line is located at $(x_0,y_0)$. The anisotropy of the vortex core is described
by the parameter $\eta$. The singularity in $\zeta_0^{(-x)}$ therefore is on the $z$ axis,
while those in $\zeta_\pm^{(-x)}$ are displaced to $(\mp x_0,y_0=0)$. Here $x_0$ is determined as the point at which $f(x_0,0)=g(x_0,0) \sin\beta$, resulting in $f(x_0,0)=\sqrt{2/(2+\cos^2\beta)}$.
The vortex configuration of Eq.~(\ref{eq;parameterfinal}) with three spatially separated vortex lines is analogous to that shown in Fig.~\ref{fig:split-core-3D}(a).

\end{document}